\documentclass[12pt]{iopart}   
\usepackage{iopams}
\usepackage{epsf}
\usepackage{euscript}
\usepackage{graphicx}

\begin{document}
\title{Nodal domains on quantum graphs}

\today

\author{Sven Gnutzmann$^{1,2}$, Uzy Smilansky$^2$ and Joachim Weber$^{2,3}$}
\address {$^2$ Institut f\"ur Theoretische Physik,\ Freie Universit\"at
  Berlin,\ Arnimallee 14\, 14195 Berlin,\ Germany}
\address {$^2$ Department of
  Physics of Complex Systems,\  The Weizmann Institute of Science,\
  Rehovot 76100,\ Israel}
\address{$^3$ Fachbereich Physik ,\ Essen,\ 45117 Essen,\ Germany.}

\ead{gnutz@physik.fu-berlin.de}

\begin{abstract}
  We consider the real eigenfunctions of the Schr\"odinger operator on
  graphs, and count their nodal domains. The number of nodal domains
  fluctuates within an interval whose size equals the number of
  bonds $B$. For well connected graphs, with incommensurate bond
  lengths, the distribution of the number of nodal domains in the
  interval mentioned above approaches a Gaussian distribution in the
  limit when the number of vertices is large. The approach to this
  limit is not simple, and we discuss it in detail. At the same time
  we define a random wave model for graphs, and compare the
  predictions of this model with analytic and numerical
  computations.
\end{abstract}


\section{Introduction- the Schr\"odinger operator on graphs}
\label{introduction}

The structure of the nodal set of wave functions reflects the type of
the underlying classical flow. This was suspected and discussed a long
time ago, \cite{miller79, Simmel, Gutzwiller, Aurich}, and returned to
the focus of current research once it was shown that not only the
morphology, but the distribution of the \emph{number} of nodal
domains, is indicative of the nature of the underlying dynamics
\cite{BGS}.  This was followed by several other studies of nodal
statistics \cite{bogoschmidt, Berry02, heller02, Msg, Foltin}, and
their relation to the random waves ensemble \cite{Berry77}.

Quantum graphs are excellent paradigms of quantum chaos \cite{KS},
and in the present work we try to check to what extent the
statistics of nodal domains in graphs follow the patterns observed
in the study of wave functions of the Schr\"odinger operators in
the typical systems (eg, billiards) where quantum chaos is often
discussed. Graphs are one dimensional systems. Their complex
features stem from two facts: \textit{i.} their topology
is different from a one-dimensional interval
(except for starlike graphs they are generally not simply connected), 
\textit{ii.}
the corresponding ``classical'' dynamics is not deterministic.
Because of the different topology, Sturm's oscillation theorem \cite{sturm}
does not apply for graphs. We shall show however, that Courant's
generalization of the oscillation theorem \cite{Courant} to higher
dimensions applies, but much more can be said about the problem.
That is, the number of nodal domains of the $n$'th eigenfunction
is generically bounded between $n$ and $n_{min}$ and an explicit
expression for $n_{min}$ will be given.

The rest of this section will be devoted to the introduction of
metric graphs and the corresponding Schr\"odinger operator. The
nodal counting problem will be defined in section \ref{general}
and some general results will be presented. The distribution of
the number of nodal domains will be discussed in section
\ref{statistics}. This distribution will be calculated for
\emph{star graphs} in section \ref{star}, and the results of these
computations will be used to derive the asymptotic distribution of
the number of nodal domains in the limit of large graphs. Some
results on counting domains on a bond in starlike graphs will be
presented in section \ref{starlike}. Finally, in section
\ref{random} we shall introduce the random wave model for graphs.
The mean and variance of the distribution of the number of nodal
domains will be computed explicitly and compared with a few
numerical results.

\medskip

A graph ${\mathcal{G}}$ consist of $V$ \emph{vertices} connected by
$B$ \emph{bonds}.  The $V\times V$ \emph{connectivity matrix} is
defined by:
\begin{equation}
  C_{i,j}= \textrm{number of bonds
    connecting the vertices $i$ and $j$}  .
\end{equation}
A graph is \emph{simple} when for all $i,\ j \ :\ C_{i,j}\in [0,1]$
(no parallel connections) and $C_{i,i}=0$ (no loops). The
\emph{valence} of a vertex is $v_i = \sum_{j=1} ^V C_{i,j}$ and the
number of bonds is $B=\frac{1}{2} \sum_{i,j=1} ^V C_{i,j} $.  We
denote the bond connecting the vertices $i$ and $j$ by $b=[i,j]$.  The
notation $[i,j]$ and the letter $b$ will be used whenever we refer to
a bond without specifying a \emph{direction}: $b=[i,j]=[j,i]$. To any
vertex $i$ we can assign the set $S^{(i)}$ of bonds which emanate from
it:
\begin{equation}
  S^{(i)} =\{ \mathrm{all\; bonds}\ [i,k]\   :\  C_{i,k}=1  \}  ;\qquad
  \# [S^{(i)}] = v_i .
  \label{eq:star}
\end{equation}

We assign the natural metric to the bonds. The position $x$ of a point
on the graph is determined by specifying on which bond $b$ it is, and
its distance $x_b$ from the vertex with the \emph{smaller} index. The
length of a bond is denoted by $L_b$ and, $0\le x_b\le L_b$.

The Schr\"odinger operator on ${\mathcal{G}}$ consists of the one
dimensional Laplacian on the bonds, which must be augmented by
boundary conditions on the vertices to guarantee that the operator
is self-adjoint. We derive the form of the boundary conditions
here, since this way we can introduce several of the concepts and
definitions to be used later on. Let $x\in {\mathcal{G}}$ and
$\Psi(x)$ a real valued and continuous function on
${\mathcal{G}}$, so that $\Psi(x)= \psi_b(x_b)$ for $x \in b$, and
$0 \le x_b \le L_b$. The functions $\psi_b(x_b)$ are real valued,
bounded with piecewise continuous first derivatives. The set of
functions $\Psi(x)$ which fulfill these conditions will be denoted
by ${\mathcal{D}} $ and they are the domain of the (positive
definite) quadratic form
\begin{equation}
  Q [\Psi] = \int_{{\mathcal{G}}} \mathrm{d} x (\nabla \Psi(x) )^2
  \equiv  \sum_{b=1}^B \int_0^{L_b}{\rm d}x_b \left(
    \frac{\mathrm{d}\psi_b}{\mathrm{d}x_b} \right)^2\ .
\end{equation}
The unique self-adjoint extension for the Schr\"odinger operator, $H$,
is determined by the Euler - Lagrange extremum principle.  The domain
of $H,\ {\mathcal{D}}_H$ consists of functions in ${\mathcal{D}}$,
with twice differentiable $\psi_b(x_b)$, which satisfy the boundary
conditions
\begin{equation}
  \forall i=1,\dots,V \ :\ \sum_{b \in S^{(i)}} \left. n_b(i)
    \frac{\mathrm{d}\psi_b}{\mathrm{d}x_b} \right |_i\ =\ 0 \ ,
  \label{eq:boundary}
\end{equation}
where the derivatives are computed at the common vertex, and $n_b(i)$
takes the value $1$ or $-1$ if the vertex $i$ is approached by taking
$x_b$ to $0$ or $L_b$, respectively. These boundary conditions are
referred to as the \emph{Neumann} boundary conditions. In the
following we shall denote by $\phi_i$ the value of $\Psi$ at the
vertex $i$.

The spectrum of the Schr\"odinger operator $H$ is discrete, non
negative and unbounded. It is computed by solving
\begin{equation}
  -\frac{\mathrm{d}^2\psi^{(n)}_b(x_b)}{\mathrm{d} x_b^2} = k_n ^2
  \psi^{(n)}_b(x_b)\ ,\ \ \ \ \ \forall b.
  \label{eq:Sch}
\end{equation}
subject to the boundary conditions (\ref{eq:boundary}).  The resulting
eigenvalues are denoted by by $k_n^2$, and they are ordered so that
$k_n\le k_m$ if $n \le m$.

For later use we quote the following  property. Let
${\mathcal{D}}_n$ denote the subspace of functions in
${\mathcal{D}}$ which are orthogonal to the first $n-1$
eigenfunctions of $H$. Then, for any non zero $\Phi\in
{\mathcal{D}}_n$
\begin{equation}
  Q [\Phi] \ge k_n^2 \int_{{\mathcal{G}}}\mathrm{d}x (\Phi^2(x)) \ .
  \label{eq:RR}
\end{equation}
Equality holds if and only if $\Phi$ is the $n$'th eigenfunction
of $H$.

It is convenient to present the solutions of (\ref {eq:Sch}) on the
bond $b=[i,j]$ ($i<j$) as
\begin{equation}
  \psi_b(x_b)= \frac{1}{\sin k L_b}\left ( \phi_j\sin  k x_b +\phi_i\sin k
    (L_b-x_b)\right).
  \label{eq:explicit}
\end{equation}
The spectrum is computed by substituting (\ref{eq:explicit}) in
(\ref{eq:boundary}), which results in a set of linear and homogeneous
equations for the $\phi_i$ with $k$ dependent coefficients
$h_{i,j}(k)$. The spectrum is obtained as the solutions of the
equation $\zeta(k)\equiv \det h(k) =0$. As will be explained shortly,
we shall assume that the lengths $L_b$ are \emph{rationally
  independent}, so that $\zeta(k)$ is an almost periodic function of
$k$. We shall also restrict our attention to simple and connected
graphs, and to avoid lengthy discussions of special cases, will assume
that the valences $v_i \ge 3$ at all the vertices (exceptions will be
stated explicitly).

\section{Nodal domains on graphs}
\label{general}

Nodal domains are connected components of ${\mathcal{G}}$ where
the wave-function has a constant sign. One cannot exclude the
possibility that eigenfunctions of the Schr\"odinger operator
vanish identically on one or several bonds. This is often the case
if the bond lengths are rationally dependent. As an example,
consider three vertices which are connected by bonds which form a
triangle. If the lengths of each of the bonds are integer
multiples of $L_t$, then there exist eigenfunctions with
eigenvalues $k_{t,n} = n \frac{2\pi}{L_t}$,for any integer $n$
which vanish on all the other bonds  \cite {Holtsam}.  To exclude
such cases, we shall discuss graphs with lengths which are
\emph{rationally
  independent (incommensurate)}.

Rational independence is not sufficient to remove completely the
possibility that wave functions vanish along one or several bonds.
To construct an example, take any graph  and choose a wave
function which has a few nodal points on it. Connect the nodal
points by bonds and take their length such that the new graph has
incommensurate lengths. The Schr\"odinger operator for the newly
constructed graph has the same eigenvalue, and a wave function
which vanishes identically on all the added bonds. This
construction is quite general, but at most, it can bring about a
negligible number of such wave functions. The reason for this is
as follows. The $n$'th  wave function on the  bond $[i,j]$
vanishes if and only if both $\phi^{(n)}_i$ and $\phi^{(n))}_j$
vanish.  The vectors $\phi^{(n)}$ are the null vector  of the {\it
quasi-periodic} matrix $h_{i,j}(k_n)$, and as $k_n$ goes over the
spectrum, they cover the sphere ergodically. Thus, the probability
that several components are exactly $0$ is vanishingly small. In
the sequel we shall ignore these non-generic cases, but bear in
mind, however, that their presence cannot be completely excluded.

The nodal domains on graphs are divided into two types:

\noindent \emph{i. interior domain} - A domain  which is
restricted to a single bond , and whose length is exactly
$\frac{\pi}{k}$.

\noindent  \emph{ii. vertex domain } - A domain which includes a
vertex, and extends to the bonds which emanate from it.

There are $V$ vertex domains, and their length $\Lambda_i$ can
take any value in the range $v_i\frac{\pi}{k} > \Lambda_i\ge 0$.
Denoting the length of the graph by ${\mathcal{L}} = \sum_{b=1}^B
L_b$, we obtain the following expression for the number of nodal
domains,
\begin{equation}
\nu_n = V + \frac{k_n}{\pi}\left({\mathcal{L}}-
  \sum_{i=1}^V\Lambda_i\right )\ .
 \label{eq:NND1}
\end{equation}
Note that the second term above is an integer, and that this
expression is correct for the generic case where the wave-function
does not vanish along entire bonds. $\nu_n$ is bounded in the interval
\begin{equation}
  \frac{k_n{\mathcal{L}}}{\pi}+V \ge   \nu_n \ge \frac{k_n{\mathcal{L}}}{\pi}+ V -2B
  \ .
  \label{eq:range1}
\end{equation}

In the limit $n \rightarrow \infty , \ \frac{k_n {\mathcal{L}}}{n
  \pi}\rightarrow 1 $. Hence, in this limit, $\frac{\nu_n}{n}
\rightarrow 1$. This observation stands intermediately between
Sturm's oscillation theorem ($\frac{\nu_n}{n} = 1$), and Pleijel's
result that $\overline{\lim} \frac{\nu_n}{n} $ is strictly smaller
than $1$ for the eigenfunctions of the Dirichlet Laplacian for
domains in $\mathbb{R}^2$ \cite{pleijel}.

An alternative expression for the number of nodal domains provides a
sharper bound on the range of variation of $\nu_n$. Denoting the
number of \emph{nodal points} on the bond $b=[i,j]$ by $\mu_n^{(b)}$,
we have
\begin{equation}
  \mu_n^{(b)} = [\! [
  \frac{k_n L_b}{\pi}]\! ]+\frac{1}{2} \left( 1-(-1)^{
      [\! [\frac{k_n L_b}{\pi} ]\! ]}
    \mathrm{sign}[\phi_i]\mathrm{sign}[\phi_j]
  \right )
  \label{eq:mu}
\end{equation}
where $[\! [ x ]\! ]$ stands for the largest integer which is smaller
than $x$, and $\phi_i, \phi_j$ are the values of the eigenfunction at
the vertices $i,j$ respectively. Thus,
\begin{equation}
  \nu_n =\sum_{b=1}^B \mu_n^{(b)} -B + V \ .
  \label{eq:NNB2}
\end{equation}
The allowed range for $\nu_n$ is  now
\begin{equation}
  \sum_{b=1}^B [\! [\frac{k_n L_b}{\pi}]\! ]+V \ge   \nu_n \ge
  \sum_{b=1}^B [\! [\frac{k_n L_b}{\pi}]\! ]+V -B \ .
  \label{eq:range2}
\end{equation}

The estimates from above can be sharpened by Courant's law
\cite{Courant} adapted for the present problem, which we shall now
state and prove.

\noindent \textit{Theorem: Let ${\mathcal{G}}$ be a simple, connected
  graph. Let $k_n^2$ be the $n$'th eigenvalue of the Schr\"odinger
  operator $H$ defined above and let $\Psi_n(x)$ be the corresponding
  real eigenfunction. Then, the number of nodal domains $\nu_n$ of
  $\Psi_n(x)$ is bounded from above by $n$, and this bound is
  optimal.}

The proof follows the method used in \cite{pleijel}. Assume that
$\nu_n > n$. Denote by $\gamma_l$, the nodal domains on
${\mathcal{G}}$, so that
$\bigcup_{l=1}^{\nu_n}\gamma_l ={\mathcal{G}}$.
Construct $n$ functions $U_l(x) \in \mathcal{D}$ in
the following way:
\begin{equation}
  U_m(x) = \left\{
    \begin{array}{ll}
      \Psi_n(x) \qquad &   \mathrm{if} \  x\in \gamma_m \\
      0 & \mathrm{otherwise}
    \end{array}
  \right .
\end{equation}
It is always possible to find $n$ real constants $a_m$ such that
$U(x)= \sum_{m=1}^n a_m U_m(x) $ is orthogonal to the first $n-1$
eigenfunctions of $H$. Hence $U(x) \in {\mathcal{D}}_n$. A simple
computation shows that
\begin{equation}
  Q[U] = k_n^2 \int_{{\mathcal{G}}} U^2(x) \ .
\end{equation}
However, $ U(x)$ is not an eigenfunction, hence the above equality
is in contradiction with the strong inequality imposed by
(\ref{eq:RR}). Thus, the assumption that $\nu_n>n$ is false.$ \ \
\Box $

In the next sections we shall try to determine how the $\nu_n$ are
distributed within their allowed range. We shall start by solving a
simpler problem, which pertains to the family of \emph{star graphs}.
A similar derivation for more complicated graphs is beyond our present
ability.  However, assuming that in the limit of large graphs, the
lengths of vertex nodal domains are independent, we shall be able to
deduce an approximate expression for the distribution of $\nu_n$ in
this limit.

\section{Nodal domain statistics on graphs}
\label{statistics}

In the previous section we observed that
$\frac{\nu_n}{n}\rightarrow 1$ as $n\rightarrow \infty$. Hence,
there is no point to use the statistics proposed in \cite {BGS}
for graphs. Rather, we shall discuss the distribution of the
quantity
\begin{equation}
\delta \nu_n=\nu_n -n
\end{equation}
which can vary in the interval $\frac{k_n \pi}{\mathcal{L}}
+V-2B-n\le \delta \nu_n \le 0$. Let $\lambda_n=\frac{k_n}
{\pi}\sum_{i=1}^V \Lambda_i$ denote the sum of the lengths of the
vertex domains measured in units of half the wave length.
Following (\ref{eq:NND1}) we find
\begin{equation}
\delta \nu_n=\left[\frac{k_n
\mathcal{L}}{\pi}+\frac{1}{2}-n\right] +V-\frac{1}{2}-\lambda_n =
\delta N(k_n)+ V-\frac{1}{2}-\lambda_n \ .
  \label{eq:NND3}
\end{equation}
The expression in the square brackets above is the deviation
$\delta N(k_n)$ of the mean spectral counting function \cite{KS}
from its actual value. Thus, the fluctuations in the number of
nodal domains stem from two sources: the spectral counting
fluctuations, and the fluctuations in the lengths of the
\emph{vertex domains} $\lambda_n$, whose distribution we shall
denote by \begin{equation}
  P(\lambda)=
  \langle \delta(\lambda-\lambda_n)\rangle \ ,
  \label{eq:sizedist}
\end{equation}
where $\langle\cdots\rangle$ indicates average over a spectral
interval of size $\Delta k$, with $\frac{\Delta k {\cal L}}{\pi}$
eigenvalues on average. In general, and especially for graphs with
small $B$, the two contributions are probably correlated. For
large graphs, however, such correlations are expected to be much
weaker. This were the case if the spectral and the eigenvector
distributions are independent, like in the relevant random matrix
ensemble (GOE). We are not able to prove this statement, and we
\emph{assume} that in the limit of large graphs the two
contributions can be treated independently. The quantities of
interest here are
\begin{equation}
  \langle \delta \nu \rangle= \left\langle\left(\frac{k_n \pi}{\mathcal{L}}+
  \frac{1}{2}-n\right)\right\rangle
  +V-\frac{1}{2}-\langle\lambda_n\rangle
  \label{eq:mean}
\end{equation}
and
\begin{equation}
   \langle \Delta \delta \nu^2\rangle \approx
  \left\langle \Delta \left(\frac{k_n \pi}{\mathcal{L}}+\frac{1}{2}-n\right)^2\right\rangle
  + \langle\Delta \lambda^2 \rangle \ .
  \label{eq:variance}
\end{equation}
The contribution of the spectral fluctuations to the mean $\langle
\delta \nu \rangle$, vanishes $\mathcal{O}\left (\frac{1}{\Delta
k}\right)$.  $\lambda_n$ is bounded to $0\le \lambda_n \le 2B$. In
the sequel, we shall provide evidence in support of the natural
expectation that  the $\lambda$ distribution is symmetric around
the point $B$, thus
\begin{equation}
  \langle \lambda\rangle=B \ .
\end{equation}
 Hence,
 \begin{equation}
  \langle\nu_n-n\rangle=-(B-V+\frac{1}{2}).
\end{equation}
This result is consistent with $\nu_n-n\le 0$ since we assumed
$v_i\ge 2$ at all the vertices.

Turning to the variances, the contribution from the spectral
counting function for general systems, and for graphs in
particular, was studied previously by various authors. We show in
Appendix (A) that
\begin{equation}
\langle \delta N(k_n)\rangle\ =  \frac{B}{6}\left (1+{\cal O}\left
(  \frac{\log B}{B}\right) \right ) \ ,
 \label{eq:dnbound}
\end{equation}
The main term in (\ref{eq:dnbound}) is a universal bound which is
valid for all incommensurate Neumann graphs. The error estimate is
valid for  well connected graphs, where the spectral statistics is
known to follow the random matrix predictions.  The rest of this
and the following sections will deal with the distribution of the
total size of the vertex domains $\lambda_n$.

\begin{figure}
  \begin{center}
    \includegraphics[width=0.45\linewidth]{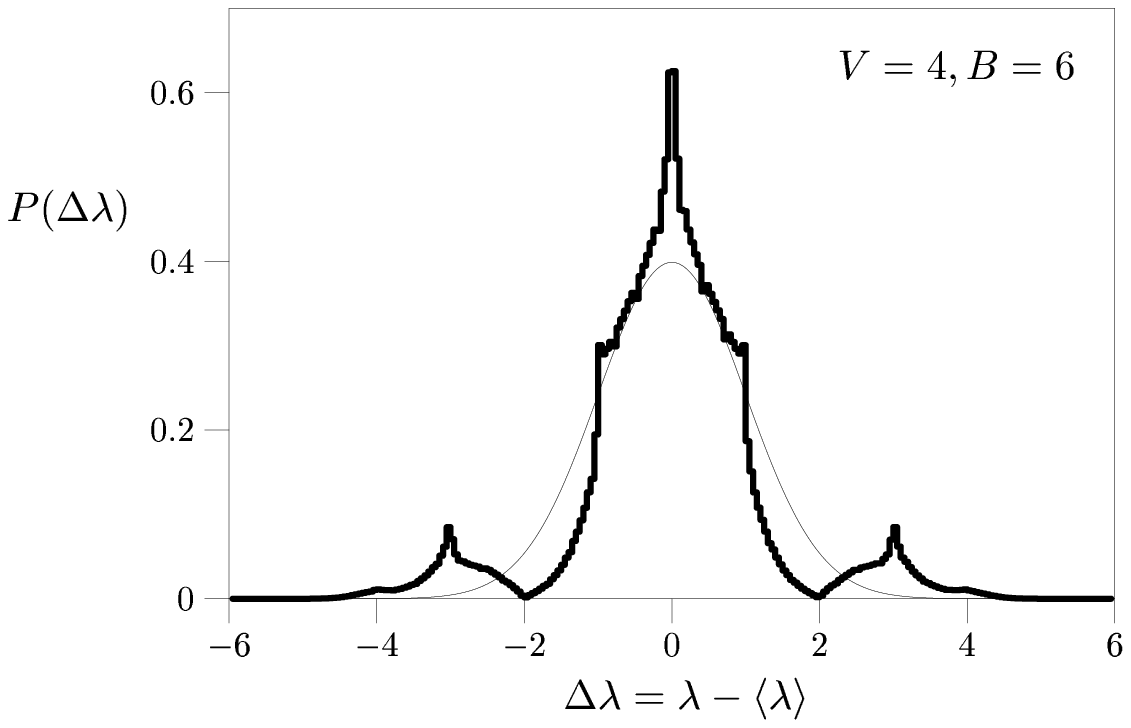}
    \includegraphics[width=0.45\linewidth]{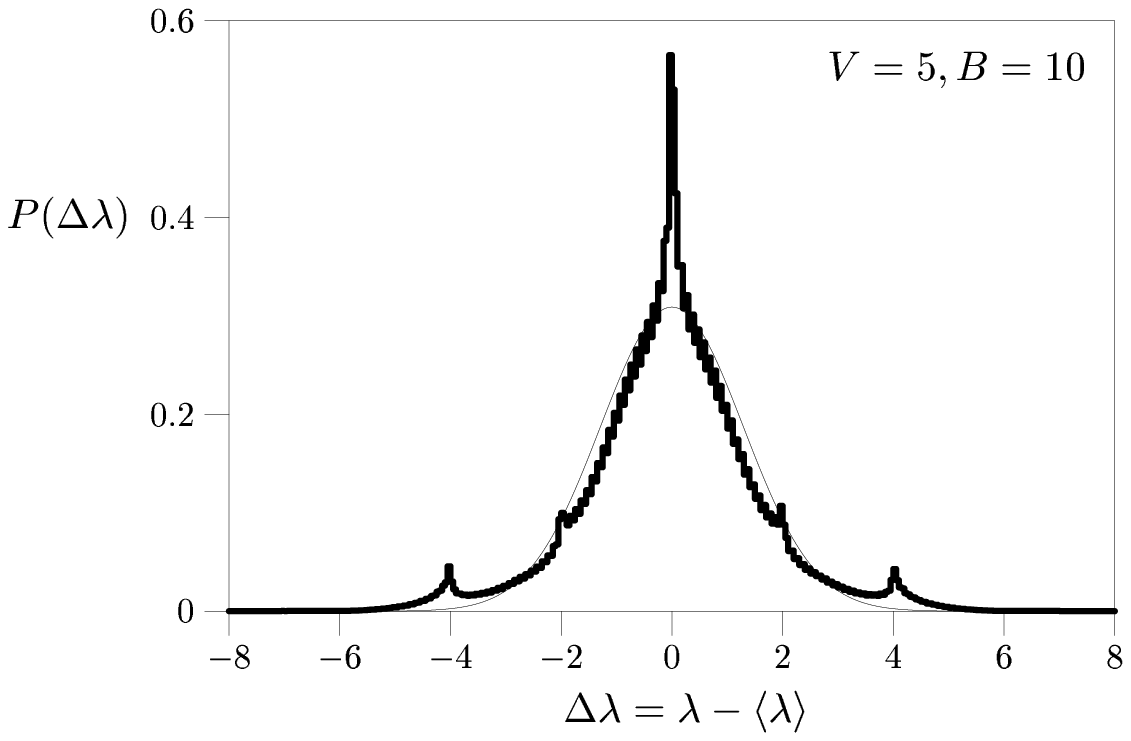}
    \includegraphics[width=0.45\linewidth]{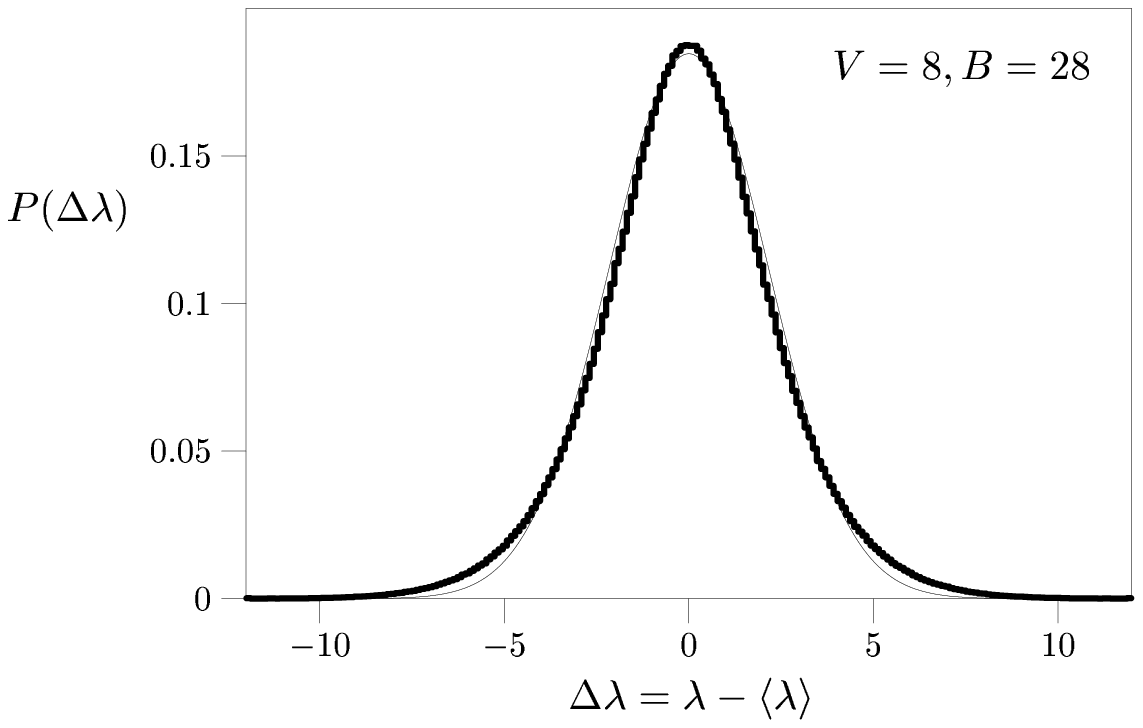}
    \includegraphics[width=0.45\linewidth]{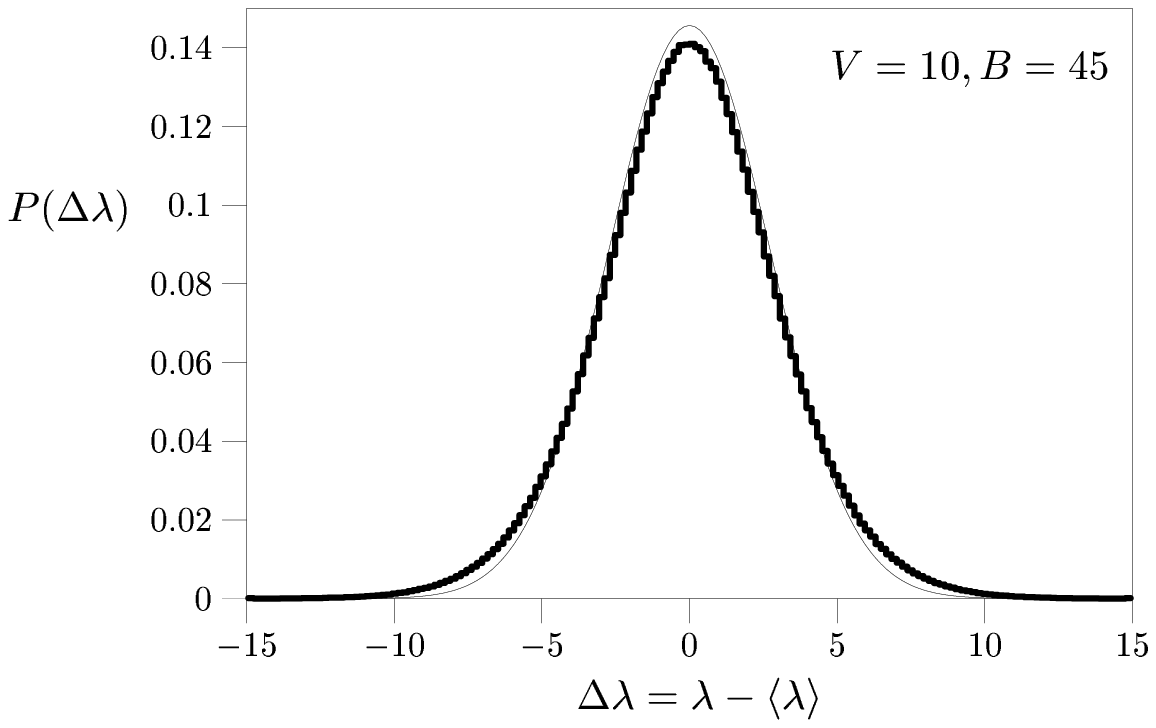}
    \caption{Distribution of the deviation $\Delta \lambda$ of the
      total length of vertex domains $\lambda$
      from its mean value $\langle \lambda \rangle=B$
      for fully connected graphs with $V=4,5,8$ and $10$ vertices
      (in units of half the wave
      length). The first $10^7$ eigenfunctions have been
      used for the numerically obtained full line.
      The thin line is a Gaussian of variance $B/6$ where
      $B=\frac{V(V-1)}{2}$.}
    \label{figure0}
  \end{center}
\end{figure}

The distribution $P(\lambda)$ for a finite small
graph shows distinctive features as can be seen in figure
(\ref{figure0}) where we plotted the numerically obtained
distribution for a fully connected graph with
$V=4$ vertices and $B=6$ bonds (the \emph{tetrahedron})
and compare it to larger graphs.
A bell shaped function is
obtained for $P(\lambda)$ of larger graphs which is quite well
approximated by a Gaussian of variance $\frac{B}{6}$.

Since $\lambda_n$ is bounded in the interval  $(0,2B)$ its
variance cannot grow faster than $B$
\begin{equation}
   \langle\Delta \lambda^2 \rangle= \beta({\cal G}) B\ ,
\end{equation}
where $\beta({\cal G})<1$. We shall compute $\beta({\cal G}) $
below for particular models, and show that in general, for large
graphs $\beta({\cal G})=\frac{1}{6}$.

A quantity which might be of some interest in the present context
is the length (again in units of half the wave length)
$\chi^{(i,j)}_n$ of the intersection of a vertex domain at a given
vertex $i$ with the single bond $b=[i,j]$. Generally,
$\chi^{(i,j)}_n \neq \chi^{(j,i)}_n$ but they are related by
\begin{equation}
  \begin{array}{ll}
    \displaystyle
    L_{[i,j]}=&\displaystyle\frac{\pi}{k_n}
    \Big([\![ \frac{k_n L_{[i,j]}}{\pi}]\!]+
    \chi^{(i,j)}_n +\chi^{(j,i)}_n \\
    &\displaystyle
    -\frac{1}{2}-\frac{(-1)^{[\![ \frac{k_n L_{[i,j]}}{\pi}]\!]}}{2}\,
    \mathrm{sign}[\phi_i]\,\mathrm{sign}[\phi_j]\Big)\ .
  \end{array}
  \label{eq:corr_chi}
\end{equation}
The total length of all vertex domains is
\begin{equation}
  \lambda_n=\sum_{i<j} C_{i,j}\left( \chi^{(i,j)}_n +\chi^{(j,i)}_n
  \right)\ .
  \label{eq:domainsize}
\end{equation}
The distribution
\begin{equation}
  P^{(i,j)}(\chi)= \langle
  \delta(\chi-\chi^{(i,j)}_n) \rangle
  \label{eq:bonddist}
\end{equation}
is thus connected to nodal counting on a single bond. Due to (strong)
correlations between the $\chi^{(i,j)}_n$ for different $i$ and $j$
(\ref{eq:bonddist})
is less useful than (\ref{eq:sizedist}) or nodal counting on a complete
graph.

\subsection{Nodal domain statistics on star graphs}
\label{star}

\begin{figure}
  \begin{center}
    \includegraphics[width=0.2\linewidth]{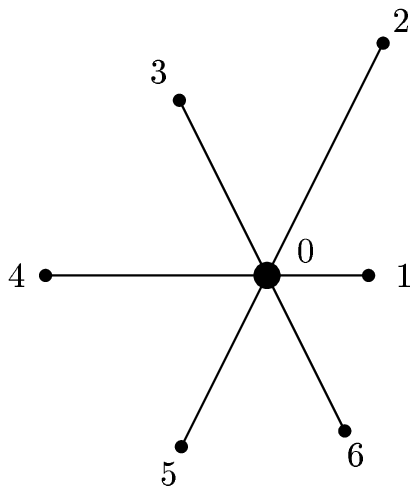}
    \caption{A star graph with $B=6$ bonds emanating from the central
      vertex $0$.}
    \label{figure1}
  \end{center}
\end{figure}

In a star graph all bonds emanate starlike from one central vertex $i=0$. Each
bond $b_i=[0,i]$ ($i=1,\dots,B$) connects the central vertex to one
\emph{peripheral vertex} $i$ (see figure \ref{figure1}). The
bond $b_i$ has the length $L_i$ and all lengths are chosen
incommensurate. The variable $x_i$ measures the distance
from the center on bond $b_i$ such that $0\le x_i \le L_i$
and $x_i=L_i$ at the peripheral vertex $i$. Though the
number of vertices is $V=B+1$ only the central vertex fulfills $v_0
\ge 3$ (if $B \ge 3$).  The peripheral vertices have valence $v_i=1$
and instead of Neumann boundary conditions we will use Dirichlet
boundary conditions $\phi_i=0$ ($i=1,\dots,B$) there.  The wave
function on the bond $b_i$ follows from (\ref{eq:explicit})
\begin{equation}
  \psi_i(x_i)= \frac{\phi_0}{\sin k L_i}\sin  k \left(L_i-x_i\right)
\end{equation}
where $\phi_0$ is the value of the wave
function on the central vertex. Current conservation at the center
leads to the quantization condition
\begin{equation}
  f_B(k_n)\equiv \sum_{i=1}^B \mathrm{cot}\ k_n L_i =0
  \label{eq:quant_star}
\end{equation}
for the $n$'th eigenvalue $k_n$ of the star graph.

Since the peripheral vertices are nodal points, equations
(\ref{eq:NND1}) and (\ref{eq:NND3}) for the number of nodal
domains have to be modified to
\begin{equation}
  \nu_n=1+\sum_{i=1}^B\, [ \! [
  \frac{L_i k_n}{\pi} ] \! ] =1+
  \frac{k_n}{\pi}\mathcal{L}-\lambda_n
  \label{eq:NND_star}
\end{equation}
where $\lambda_n= \frac{k_n\mathcal{L}}{\pi}-\sum_{i=1}^B [ \! [
\frac{L_i k_n}{\pi} ] \! ]$ is $\frac{k_n}{\pi}$ times the length of
the nodal domain that contains the central vertex.
Obviously $0\le
\lambda_n \le B $ and $\nu_n$ is bounded by
\begin{equation}
  1+\frac{k_n\mathcal{L}}{\pi}\ge \nu_n \ge
  1-B+\frac{k_n\mathcal{L}}{\pi}.
\end{equation}

\subsubsection{The central nodal domains} \label{sec:central}

 $ \\ $
 As discussed above, nodal counting is partly
determined by spectral fluctuations and partly by the distribution
(\ref{eq:sizedist}) of the length $\lambda$ of the central nodal
domain. We shall consider here only the distribution of the
lengths of the central vertex domain,
\begin{equation}
  \begin{array}{ll}
    \displaystyle P(\lambda)&\displaystyle
    = \langle
    \delta(\lambda-\lambda_n)\rangle \\
    & \displaystyle =\lim_{\Delta k\rightarrow \infty}
    \frac{\pi}{\Delta k \mathcal{L}}
    \int_0^{\Delta k} \mathrm{d}k\, \left|\frac{\mathrm{d}f_B}{\mathrm{d}k} (k)
    \right|\delta(f_B(k))  \delta(\lambda-\lambda(k))
  \end{array}
  \label{eq:dist1}
\end{equation}
where $\lambda(k)=\sum_{i=1}^B \left(\frac{k L_i}{\pi}- [\![
\frac{k L_i}{\pi} ]\!]\right)$. From (\ref{eq:quant_star}) we have
\begin{equation}
  \frac{\mathrm{d} f_B}{\mathrm{d} k}(k)=-\sum_{i=1}^B
  \frac{L_i}{\sin^2 k L_i} \le 0.
\end{equation}

Let
\begin{equation}
  \chi_i=\frac{k L_i}{\pi}- [ \![ \frac{k L_i}{\pi} ]\! ]
  \label{eq:chi_i}
\end{equation}
be the (rescaled) length of the intersection of the central nodal
domain with the $i$'th bond ($\lambda(k)=\sum_{i=1}^B \chi_i$).
Obviously, $0\le \chi_i \le 1$ and since the length $L_i$ are
assumed incommensurate, $k$ creates an ergodic flow on the
$B$-torus spanned by the $\chi_i$ \cite{Barra}. Thus, the spectral
integral over $k$ in (\ref{eq:dist1}) may be replaced by an
integral over the $B$-torus variables $\chi_i$. This leads to
\begin{equation}
\label{eq:poflambda}
  P(\lambda)=\pi \int_0^1 \mathrm{d}^B \chi\, \frac{1}{\sin^2 \pi\chi_1}\,
  \delta\!\left(\sum_{i=1}^B \mathrm{cot} \pi \chi_i\right)\,
  \delta\!\left(\lambda-\sum_{i=1}^B \chi_i\right).
\end{equation}
Replacing the two $\delta$-functions by their Fourier
representation, the distribution takes the form
\begin{equation}
  P(\lambda)=\frac{1}{4\pi}\int_{-\infty}^\infty \mathrm{d} \eta\,
  \int_{-\infty}^\infty \mathrm{d} \xi\,
  G(\eta,\xi)^{B-1}\tilde{G}(\eta,\xi)\,
  \mathrm{e}^{\mathrm{i}\xi(\lambda-\frac{B}{2})}
\end{equation}
where
\begin{equation}
  G(\eta,\xi)=\frac{2}{\pi}\int_0^{\frac{\pi}{2}} \mathrm{d} \alpha
  \, \cos \left(\eta \tan \alpha +\frac{\xi}{\pi}\alpha\right)
  \label{eq:G}
\end{equation}
and
\begin{equation}
\label{eq:G-tilde}
  \begin{array}{ll}
    \displaystyle
    \tilde{G}(\eta,\xi)&\displaystyle
    =\left(1-\frac{\partial^2}{\partial \eta ^2}\right)
    G(\eta,\xi)\\
    &\displaystyle =2\, \cos \frac{\xi}{2}\,\delta(\eta) - \frac{\xi}{\pi} \,
    G(\eta,\xi)\, \mathbb{P} \frac{1}{\eta}\ .
  \end{array}
\end{equation}
The last line shows, that $\tilde{G}(\eta,\xi)$ is a distribution
where $\mathbb{P} \frac{1}{\eta}$ denotes the Cauchy's principal
value. The integral (\ref{eq:G}) can be solved explicitly (see
\cite{Gradsteyn}, (3.718)) in terms of Whittaker functions
$W_{\mu,\nu}(x)$
\begin{equation}
  G(\eta,\xi)=\Theta(\eta)
  \frac{W_{-\frac{\xi}{2\pi},\frac{1}{2}}(2\eta)}{\Gamma(1-\frac{\xi}{2\pi})}
  +\Theta(-\eta)
  \frac{W_{\frac{\xi}{2\pi},\frac{1}{2}}(-2\eta)}{\Gamma(1+\frac{\xi}{2\pi})}.
\end{equation}
Here $\Theta(x)$ is Heaviside's step function. The appearance of
Whittaker functions can also be seen from (\ref{eq:G-tilde}) --
for $\eta \neq 0$ the right hand sides reduce to
$(1-\frac{\partial^2}{\partial \eta ^2}) \, G(\eta,\xi)=
-\frac{\xi}{\pi \eta} \, G(\eta,\xi)$, a special case of
Whittaker's differential equation \cite{Gradsteyn}. Using the last
line of equation (\ref{eq:G-tilde}) the distribution can be written
as a sum of two terms
$P(\lambda)=P_{\delta}(\lambda)+P_{\mathbb{P}}(\lambda)$ where
\begin{equation}
  \begin{array}{lll}
    \displaystyle P_{\delta}(\lambda)&\displaystyle
    =&\displaystyle \frac{1}{2\pi}\int_{-\infty}^{\infty} \mathrm{d} \xi
    \int_{-\infty}^{\infty} \mathrm{d} \eta\, G(\eta,\xi)^{B-1}\,
    \delta(\eta) \cos \frac{\xi}{2} \cos (\xi(\lambda-\frac{B}{2}))\\
    &\displaystyle =&\displaystyle
    \frac{2}{\pi} \int_0^\infty \mathrm{d} z\,
    \left(\frac{\sin z}{z}\right)^{B-1}
    \cos z\, \cos (z(2\lambda-B)) \\
    &\displaystyle = &\displaystyle \frac{\scriptstyle B-1}{\scriptstyle 2}
    \sum_{0 \le l < \frac{B}{2}} \frac{\scriptstyle (-1)^l}{
      \scriptstyle l!(B-1-l)!} \left(
      \Theta \left({\scriptstyle \frac{B}{2}-l-|\lambda-\frac{B}{2}|}\right)
      \left({\scriptstyle \frac{B}{2}-l-|\lambda-\frac{B}{2}|}
      \right)^{\scriptscriptstyle B-2}\right.
      +\\
      && \qquad \left.
      \Theta \left({\scriptstyle \frac{B}{2}-1-l-|\lambda-\frac{B}{2}|}\right)
      \left({\scriptstyle \frac{B}{2}-1-l-
          |\lambda-\frac{B}{2}|}\right)^{\scriptscriptstyle B-2}
    \right)
  \end{array}
  \label{eq:P-delta}
\end{equation}
and
\begin{equation}
  \begin{array}{lll}
    \displaystyle P_{\mathbb{P}}(\lambda)&\displaystyle =&
    \displaystyle -
    \frac{1}{4\pi^2}\int_{-\infty}^{\infty} \mathrm{d} \xi\, \xi
    \cos (\xi(\lambda-\frac{B}{2}))
    \int_{-\infty}^{\infty} \mathrm{d} \eta\, \mathbb{P}\frac{1}{\eta}\,
    G(\eta,\xi)^{B} \\
    &\displaystyle =&
    \displaystyle \frac{2}{\pi^2}\,\int_{0}^{\infty} \mathrm{d} z \, z
    \cos (z( 2 \lambda-B)) \times \\
    &&\displaystyle
    \int_{0}^{\infty} \mathrm{d} y\, \frac{1}{y}\,\left(
      \left(
        \frac{W_{\frac{z}{\pi},\frac{1}{2}}(y)}{
          \Gamma(1+\frac{z}{\pi})}
      \right)^B-
      \left(
        \frac{W_{-\frac{z}{\pi},\frac{1}{2}}(y)}{
          \Gamma(1-\frac{z}{\pi})}
      \right)^B
    \right) \ .
  \end{array}
  \label{eq:P-principal}
\end{equation}
$P(\lambda)$ is symmetric in $\lambda-\frac{B}{2}$. Hence $\langle
\lambda \rangle = \frac{B}{2}$.

\begin{figure}
  \begin{center}
    \includegraphics[width=0.4\linewidth]{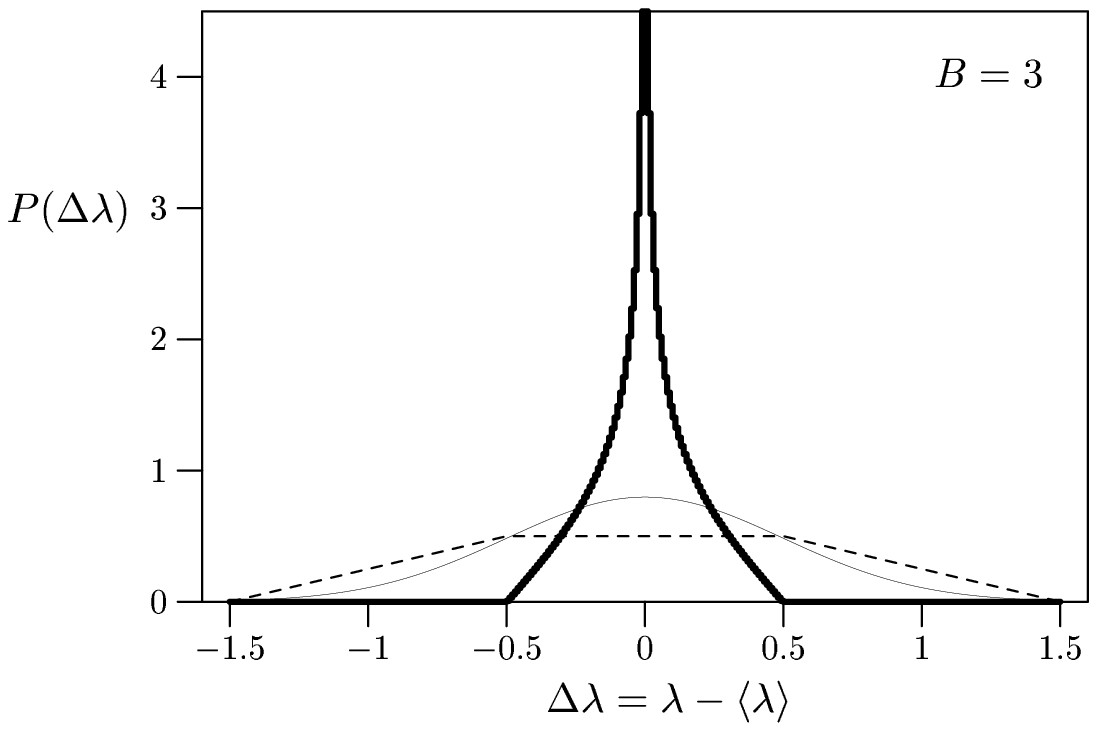}
    \includegraphics[width=0.4\linewidth]{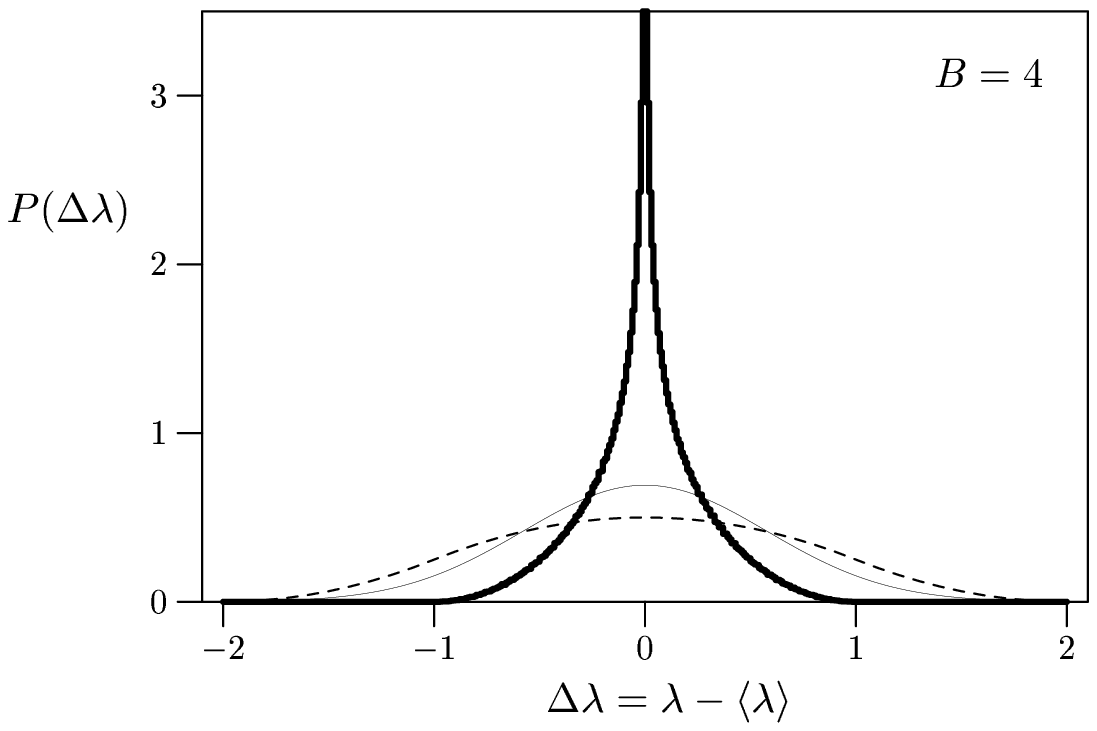}\\
    \includegraphics[width=0.4\linewidth]{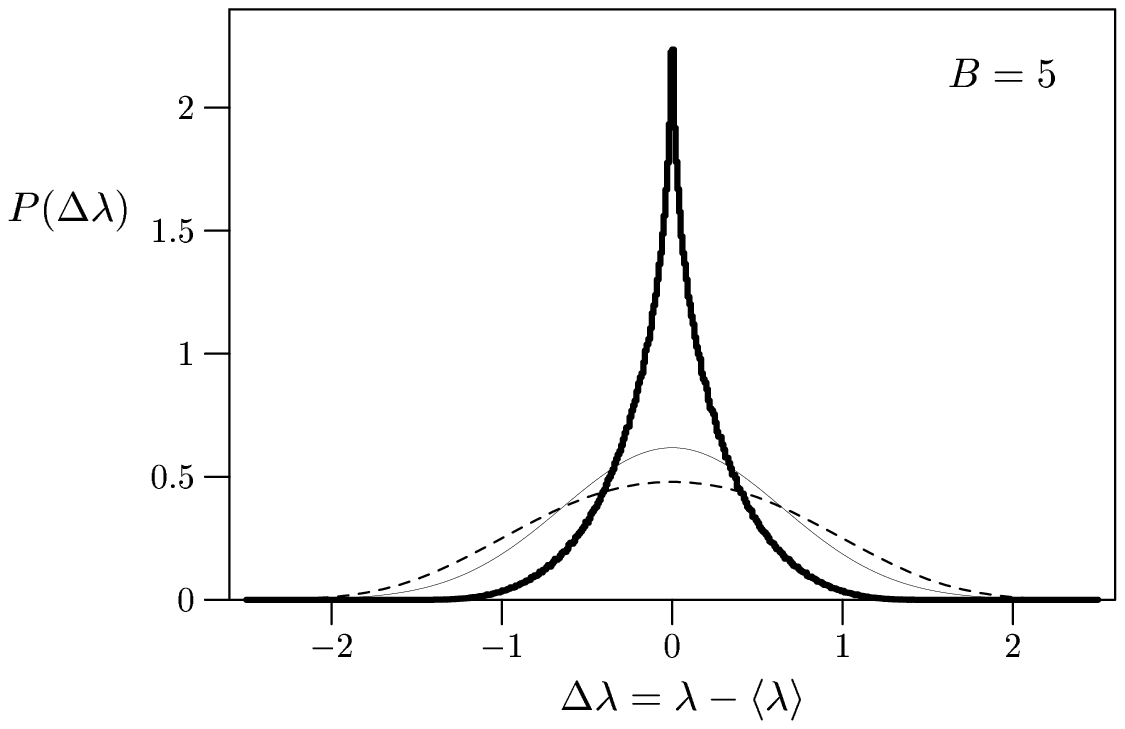}
    \includegraphics[width=0.4\linewidth]{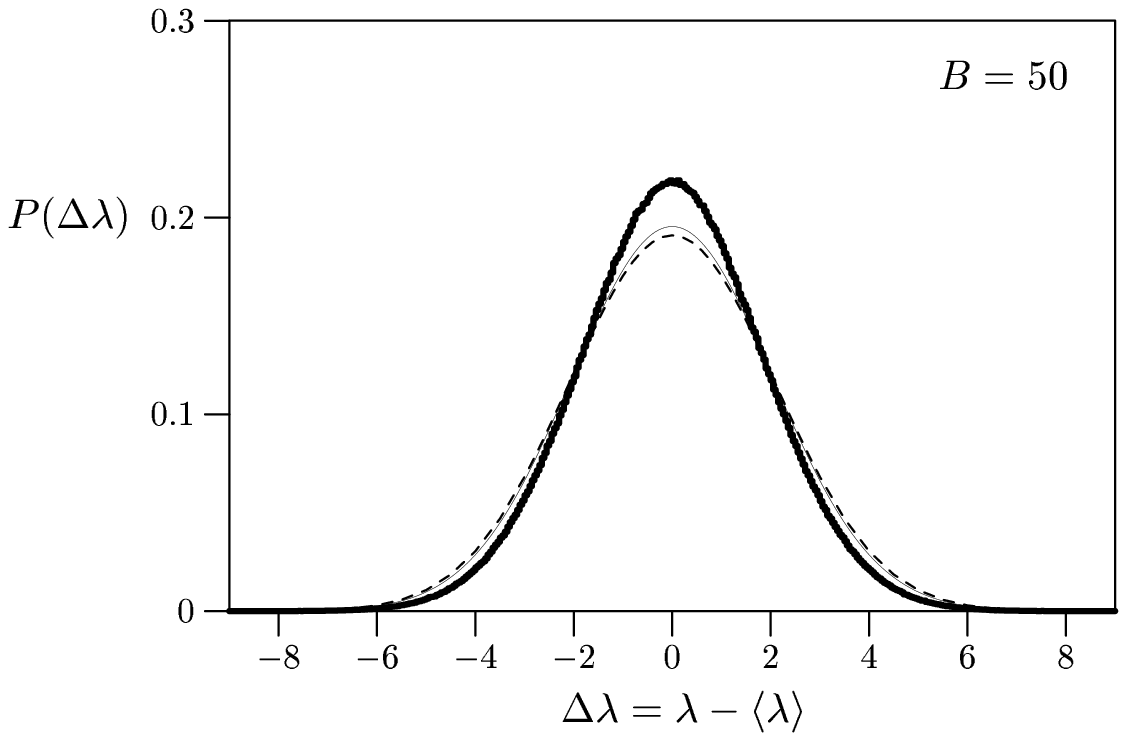}\\
    \includegraphics[width=0.4\linewidth]{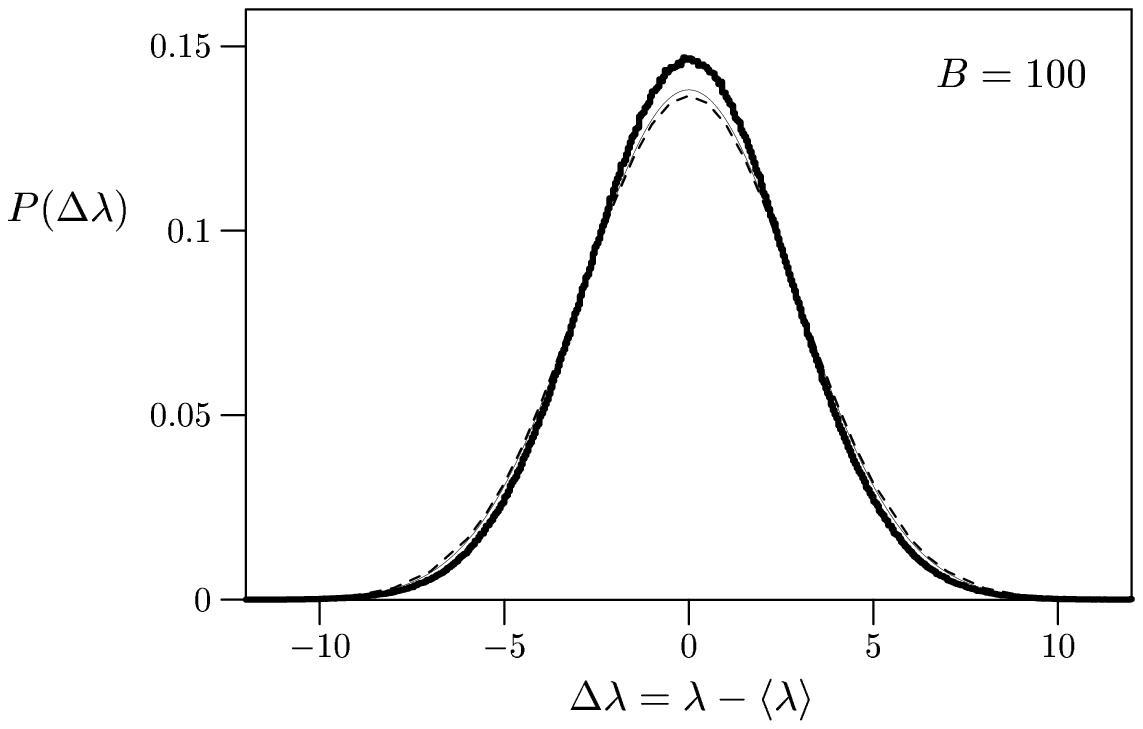}
    \includegraphics[width=0.4\linewidth]{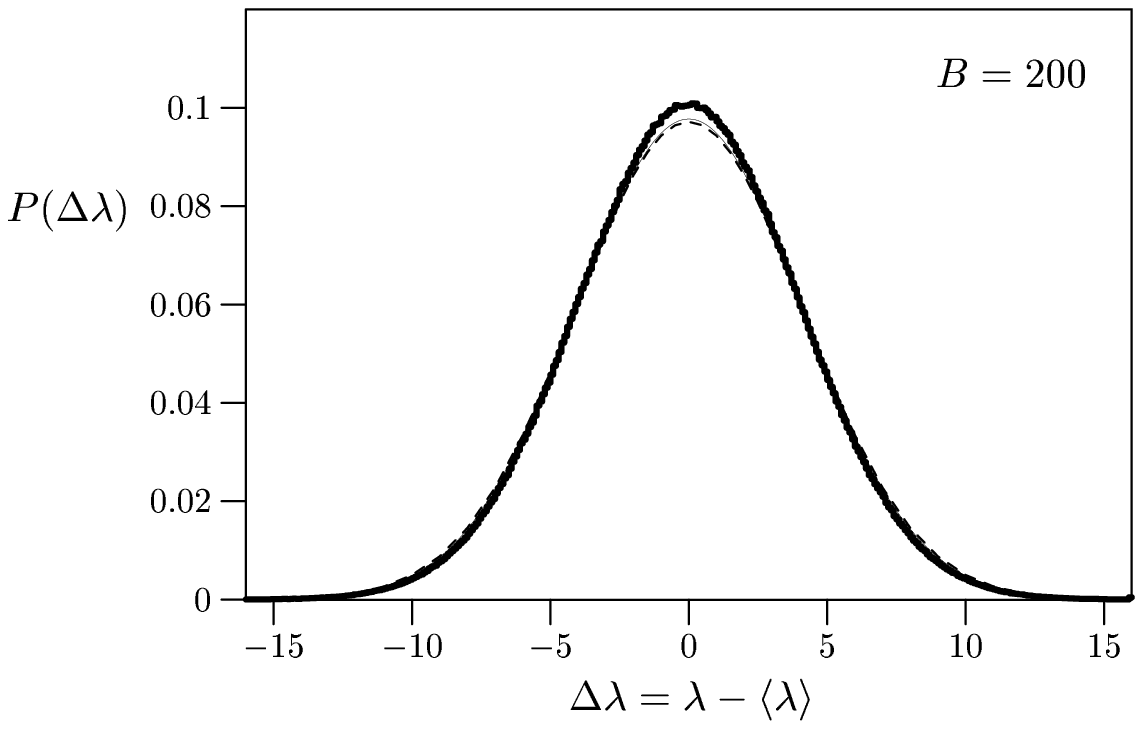}\\
    \caption{Numerically obtained distributions of
      the length of the central nodal domain in a star graph
      with $B=3$, $4$, $5$, $50$, $100$ and $200$ bonds
      (histograms -- full lines) -- the first $2\cdot 10^6$
      eigenfunctions have been used for each graph. The dashed line
      gives $P_{\delta}(\lambda)=P(\lambda)-P_{\mathbb{P}}(\lambda)$
      and the thin line is a Gaussian with variance $\frac{B}{12}$ --
      for large $B$ the Gaussian is indistinguishable from the
      $P_{\delta}(\lambda)$ and the numerical distribution slowly
      converges to the Gaussian.
      }
    \label{figure2}
  \end{center}
\end{figure}

For large star graphs ($B\gg1$), $P(\lambda)$ is dominated by
$P_{\delta}(\lambda)$ (see figure \ref{figure2}). This observation
is supported by the fact that $P_{\mathbb{P}}(\lambda)$ does not
contribute to the integrated probability distribution,
$\int_{-\infty}^{\infty} \mathrm{d}\lambda\,
P_{\mathbb{P}}(\lambda)=0$ while $\int_{-\infty}^{\infty}
\mathrm{d}\lambda\, P_{\delta}(\lambda)=1$. The dominance of
$P_{\delta}(\lambda)$ can be further supported by computing  the
variance of the exact distribution $P(\lambda)$ and of
$P_{\delta}(\lambda)$. The exact variance is evaluated by going
back to (\ref{eq:poflambda}):
\begin{equation}
  \begin{array}{ll}
    \displaystyle  \langle\Delta \lambda^2\rangle&\displaystyle
    =\int \mathrm{d}\lambda \,
    \left(\lambda-\frac{B}{2}\right)^2 P(\lambda)\\
    &\displaystyle
   =\pi \int_0^1 \mathrm{d}^B \chi\, \frac{1}{\sin^2 \pi\chi_1}\,
  \delta\!\left(\sum_{i=1}^B \mathrm{cot} \pi \chi_i\right)\,
   \left(\frac{B}{2}-\sum_{i=1}^B \chi_i\right)^2
    \\
    &\displaystyle
    =\frac{B+2}{12}-\frac{4}{\pi}\int_0^{\frac{1}{2}} \mathrm{d} z
    \, z \, \mathrm{arctan} \, \frac{\tan z \pi}{B-1}\\
    &\displaystyle =\frac{B+2}{12}+ \mathcal{O}(B^{-1-\rho})\ \ ,\  \ (\rho >0) .
  \end{array} \ ,
\end{equation}
Using (\ref {eq:P-delta}) we reproduce the leading terms in the
exact variance:
\begin{equation}
\label{eq:vardelta}
  \int \mathrm{d}\lambda \,
  \left(\lambda-\frac{B}{2}\right)^2 P_{\delta}(\lambda)
  =\frac{B+2}{12}.
\end{equation}
The fact that $P_{\delta}(\lambda)$ approaches  $P (\lambda)$ for
large star graphs is very important in the present context. First,
it provides an analytic expression, which for large $B$ tends to a
Gaussian with a variance given by (\ref {eq:vardelta}). Second,
when we consider general large graphs, the size of the vertex
domains become statistically independent, and their distribution
can be approximated by a Gaussian whose variance is
\begin{equation}
\label{eq:estvar} \langle (\Delta \lambda)^2\rangle \approx
\frac{1}{12} \sum v_i = \frac{B}{6} \ \ , \ \ {\rm hence}\ \
\beta({\cal G})=\frac{1}{6}
\end{equation}
where here $B$ stands for the number of bonds on the general
graph. This result is consistent with the numerical data
shown in figure \ref{figure0}.

Combining the two estimates for the variances of the spectral
fluctuations (\ref {eq:dnbound}) and the nodal domain fluctuations
(\ref {eq:estvar}), we obtain the leading term for the variance of
the number of nodal domains:
\begin{equation}
 \langle (\Delta \nu)^2 \rangle = \frac{B}{3}.
 \label{eq:varnu}
\end{equation}
This estimate holds in the limit of large graphs. In the next
section we shall show that the random wave model for the graph
provides the same answer for the variance of the
nodal domain distribution.

\subsubsection{Nodal domains on a single bond }
 \label{starlike}

$ \\ $ In a star graph the number of nodal domains on the bond
$b_i=[0,i]$ is
\begin{equation}
   [\![ \frac{k_n L_i}{\pi}]\!]=\frac{k_n L_i}{\pi} -\chi_i
\end{equation}
where $\chi_i$ is the length of the intersection of the central
nodal domain with the bond $b_i$ as in equation (\ref{eq:chi_i})
above. Note, that there are no vertex domains on the peripheral
vertices and equation (\ref{eq:corr_chi}) has to be modified.
Following the preceding section we define
\begin{equation}
  \begin{array}{ll}
    \displaystyle P^{(i)}(\chi)&\displaystyle =\lim_{\Delta k\rightarrow \infty}
    \frac{\pi}{\Delta k \mathcal{L}}
    \int_0^{\Delta k} \mathrm{d}k\, \left|\frac{\mathrm{d}f_B}{\mathrm{d}k} (k)
    \right|\delta(f_B(k))  \delta(\chi-\chi_i(k))\\
    &\displaystyle
    =\pi \int_0^1 \mathrm{d}^B \chi\,
    \delta\!\left(\sum_{i=1}^B \mathrm{cot} \pi \chi_i\right)\,
    \delta\!\left(\chi-\chi_i\right)\sum_{j=1}^B
    \frac{L_j}{\mathcal{L}\sin^2 \pi\chi_j}\ .
  \end{array}
\end{equation}
With similar techniques as in the previous section this integral
can be solved explicitly
\begin{equation}
  P^{(i)}(\chi)=\frac{\mathcal{L}-L_i}{\mathcal{L}}+
  \frac{L_i}{\mathcal{L}} \frac{B-1}{(B-1)^2 \sin^2 \chi\pi +
    \cos^2 \chi \pi}\ .
  \label{eq:bonddist-explicit}
\end{equation}
For large star graphs $B\gg 1$ where each bond length
is of similar size, one has $\frac{L_i}{\mathcal{L}} \sim
\frac{1}{B}$, and the distribution becomes uniform (with two
singular points at $\chi=0$ and $\chi=1$). If one bond length
$L_i$ exceeds the other bond lengths such that $L_i \gg \mathcal{L}-L_i$
the distribution becomes
\begin{equation}
  P^{(i)}(\chi) \approx \frac{B-1}{(B-1)^2 \sin^2 \chi\pi +\cos^2 \chi \pi}
\end{equation}
which for large $B$ is peaked at $\chi=0$ and $\chi=\pi$.

\section{Random waves on graphs}
\label{random}

Since Berry's seminal work \cite{Berry77} random waves
have been a paradigm for chaotic wave functions. Recently
they have been used extensively in the investigation of the
nodal structure in chaotic wave functions
\cite{BGS,bogoschmidt, Berry02, heller02, Msg, Foltin}.

In this section we will introduce random waves on graphs. Any
ensemble of random waves should solve Schr\"odinger's equation on
the bonds and be continuous at the vertices. Thus, any set of
values $\left\{\phi_j\right\}$ ($j=1,\dots,V$) for the wave
function on the vertices determines a wave function on the graph
which solves  (\ref{eq:explicit}). However, these  waves do not
fulfill the correct boundary conditions (current conservation) on
the graph. The ensemble of random waves on a graph is therefore
defined in terms of the probability distribution of the $\phi_j$.

Before we define the appropriate ensemble for
more general graphs it will be instructive to
discuss star graphs. If we want to compare random waves with
the star graph results of the previous chapter we have to
keep the Dirichlet boundary conditions at the peripheral vertices
and only relax the the current conservation condition at the
center. Then $\psi_i(x_i)=\frac{\phi_0}{\sin k L_i} \sin k (L_i-x_i)$
is the random wave on the $i$'th bond and $\phi_0$ is the only
random parameter for fixed $k$. Obviously the position of nodal
points does not depend on $\phi_0$. Thus, we do not need the
distribution $P(\phi_0)$ for the discussion of nodal domain
statistics. Let us now rewrite equation (\ref{eq:NND_star})
for the number of nodal domains
\begin{equation}
  \begin{array}{ll}
    \displaystyle
    \nu(k)& \displaystyle
    =1+\sum_{i=1}^B \, \int_{1/2}^{L_i} \mathrm{d} x_i
    \sum_{j=-\infty}^{\infty} \delta(x_i - j)\\
    &\displaystyle
    =1+\frac{k\mathcal{L}}{\pi}-\frac{B}{2}+\sum_{i=1}^B
    \sum_{j=1}^{\infty}
    \frac{\sin 2 \pi j L_i k}{j \pi} \ .
  \end{array}
\end{equation}
Averaging over a $k$-interval reveals that the mean is
$\langle\nu\rangle=1+\frac{k\mathcal{L}}{\pi}-\frac{B}{2}$ --
consistent with the result for the eigenfunctions in the previous
chapter. For the variance we get
\begin{equation}
  \begin{array}{ll}
    \displaystyle \langle\Delta \nu^2\rangle & \displaystyle
    = \int_{k_0}^{k_0+\Delta k}
    \sum_{i,i'=1}^B \sum_{j,j'=1}^\infty
    \frac{\scriptstyle \cos2 \pi k(j L_i-j' L_{i'}) -
      \cos2 \pi k(j L_i+ j' L_{i'}) }{\scriptstyle
      2\, \Delta k \, j j' \,\pi^2}\\
     & \displaystyle
    =\frac{B}{2 \pi^2}\sum_{j=1}^\infty \frac{1}{j^2}
    +\mathcal{O}(\Delta k^{-1})
    = \frac{B}{12}+\mathcal{O}(\Delta k^{-1})
  \end{array}
\end{equation}
which again coincides with the result for the eigenfunctions of
star graphs for large $B$. Note, that for incommensurate bond
lengths there are no correlations between the contributions from
single bonds. Thus the number of nodal domains is a sum of
independent quantities of finite variance each. The central limit
theorem leads to Gaussian statistics for large $B$.

To define an appropriate ensemble for more general graphs we will
be guided by wave functions that \emph{do} fulfill current
conservation (we will again assume that each vertex has a valence
$v_i\ge 3$). The quantization condition for a graph has the form
$\det h_{ij}(k)=0$ where $h_{ij}(k)$ is a real symmetric matrix of
dimension $V \times V$ \cite{KS}. If the quantization condition is
fulfilled for $k$, the eigenvector for the zero eigenvalue is a
set of vertex values $\left\{\phi_j\right\}$ that determines the
eigenfunction. For incommensurate lengths $h_{ij}(k)$ is a
quasi-periodic function of $k$ such that the matrices $h_{ij}(k)$
are expected to be typical members of the Gaussian orthogonal
ensemble (GOE) in random-matrix theory. Since for that ensemble
eigenvectors have uncorrelated components, we will \emph{assume}
for the ensemble of random waves on the graph that the $\phi_j$
are independent Gaussian variables of equal variance. From
equation (\ref{eq:mu}) and (\ref{eq:NNB2}) we see that the number
of nodal domains only depends on the signs of $\phi_j$. Let
$\sigma_j=\mathrm{sign}[\phi_j]$, then $\sigma_j=\pm 1$ with equal
probability. The number of nodal domains can now be rewritten as
\begin{equation}
  \nu(k)=\frac{\textstyle k \mathcal{L}}{\textstyle \pi}
  +{\textstyle V}-{\textstyle B}+\sum_{i,j:i<j} {\textstyle C_{i,j}}\left(
    \sum_{m=1}^\infty
    \frac{\scriptstyle \sin 2\pi m k L_{ij}}{\scriptstyle m \pi}  -
    \frac{\textstyle 1}{\textstyle 2}({\textstyle -1})^{[\![
      \frac{\scriptscriptstyle k L_{ij}}{\scriptscriptstyle \pi}]\!]}
    {\textstyle \sigma_i \sigma_j}
  \right)\ .
  \label{eq:noddom_rndwav}
\end{equation}
Averaging over a $k$-interval of length $\Delta k$ centered at
$k_0$ and over $\sigma_j$ gives the mean number of nodal domains
\begin{equation}
  \langle\nu\rangle=\frac{k_0 \mathcal{L}}{\pi}+V-B +
  \mathcal{O}(\Delta k^{-1})\ .
\end{equation}
The variance of the number of nodal domains is purely due to the
sum over $i,j$ in (\ref{eq:noddom_rndwav}).  To leading order, the
sum over sines and the sum over the signs $\sigma_i$ give
independent contributions to the variance. We have already
calculated the first part
$\langle\nu^2\rangle_{\sin}=\frac{B}{12}$ above in our discussion
of random waves on star graphs. The fluctuations due to the signs
are stronger, and they provide to the variance a term
$\langle\nu^2\rangle_{\sigma}=\frac{B}{4}$.  Hence the random wave
model predicts the variance
\begin{equation}
  \langle\nu^2\rangle=\frac{B}{3}+\mathcal{O}(\Delta k^{-1}) \ .
\end{equation}
This result reproduces the estimate (\ref {eq:varnu}), which was
derived under very different assumptions. Finally we would like to
note that bonds do not contribute independently to the number of
nodal domains. However, these correlations are not expressed in
the variance. They do contribute to the higher moments.

\section{Acknowledgments}
This research was supported by the Minerva center for Nonlinear
Physics, the Minerva foundation and a grant from the Israel
Science Foundation. SG acknowledges the kind hospitality of
the Weizmann Institute of science.

\section{Appendix A}
 In this appendix we justify the bound (\ref {eq:dnbound}) on the
variance of the spectral counting function for graphs.

The starting point is the expression of the  spectral  counting
function $N(k)$  as a sum of its mean value (Weyl's law) and an
oscillatory part,
\begin{equation}
  N(k )=\frac{k  \mathcal{L}}{\pi} +\frac{1}{2}+\delta N(k),
\end{equation}
and the oscillatory part is given by \cite{KS}
\begin{equation}
  \delta N(k )=\frac{1}{\pi}\mathrm{Im}\, \sum_{m=1}^\infty \frac{
    \mathrm{tr}\, \left( \mathcal{S}_{\mathrm{B}}(k )\right)^m}{m}\
    .
  \label{eq:trace}
\end{equation}
$\mathcal{S}_{\mathrm{B}}(k_n)$ is the \emph{bond scattering
matrix} defined in \cite{KS}.  $\mathrm{tr}\, \left(
\mathcal{S}_{\mathrm{B}}(k )\right)^m$, is a sum of contributions
from all the $m-$periodic orbits on the graph.  Each contribution
is  endowed with a phase proportional to $k l^{m}_p$, where
$l^{m}_p$ is the length of the orbit, and $p$ is the summation
index. Because of the rational independence of the bond lengths,
\begin{equation}
\langle\mathrm{tr}\, \left( \mathcal{S}_{\mathrm{B}}(k )\right)^m
\mathrm{tr}\, \left( \mathcal{S}^*_{\mathrm{B}}(k
)\right)^n\rangle = 2B K_m \delta_{m,n} + \mathcal{O}(\Delta
k^{-1})
\end{equation}
Where $K_m$ is the spectral form factor associated with the graph.
The mere fact that the bond lengths of the graph are not
commensurate, is enough to guarantee that $K_m \le 1$.
\begin{equation}
\langle\left(\delta N(k )\right)^2\rangle \approx
\frac{1}{2\pi^2}\sum_{m=1}^{\infty}\frac{\langle | \mathrm{tr}\,
\left( \mathcal{S}_{\mathrm{B}}(k )\right)^m|^2\rangle}{m^2} \le
\frac{B}{6}.
\end{equation}
A sharper estimate can be given for well connected graphs, where
numerical and analytic results \cite {KS,BSW} show that the $K_m$
follow the predictions of random matrix theory for the circular
orthogonal ensemble (COE). We can use  the known functional
dependence of $K_m$ on $m$ and $B$ \cite {Fritz} and show that
\begin{equation}
\langle\left(\delta N(k)\right)^2\rangle =\frac{B}{6}(1+ {\cal O}
\frac{\log B}{B}).
 \end{equation}

\end{document}